\documentclass{article}
\usepackage[dvips]{color}
\usepackage{amsmath}
\usepackage{amsfonts}   
\usepackage{amssymb}    
\usepackage{multicol}
\usepackage{mathptm,graphicx}
\usepackage{epsf}
\newtheorem{theorem}{Theorem}
\newtheorem{lemma}{Lemma}
\newtheorem{remark}{Remark}
\newtheorem{definition}{Definition}
\newtheorem{example}{Example}
\newtheorem{proposition}{Proposition}

\begin{document}
\title{G-consistent price systems and bid-ask pricing
for European contingent claims under Knightian uncertainty}
\author{Wei Chen \\
Institute of Quantitative Economics\\
 School of Economics\\
    Shandong University\\
    250199, Jinan, China\\
weichen@sdu.edu.cn  }
\date{}
\maketitle
\begin{center}
\begin{minipage}{120mm}
\baselineskip 0.2in {\small {\bf Abstract} The target of this
paper is to consider model the risky asset price on the financial
market under the Knightian uncertainty, and pricing the ask
(upper) and bid (lower) prices of the uncertain risk. We use the
nonlinear analysis tool, i.e., G-frame work \cite{PengC}, to
construct the model of the risky asset price and bid-ask pricing
for the European contingent claims under Knightian uncertain
financial market.

\ \ \ \ First, we consider the basic risky asset price model on
the uncertain financial market, which we construct here is the
model with drift uncertain and volatility uncertain. We describe
such model by using generalized G-Brownian motion and call it as
G-asset price system. We present the uncertain risk premium which
is uncertain and distributed with maximum distribution
$N([\underline{\mu},\overline{\mu}],\{0\})$. Under G-frame work we
construct G-martingale time consistent dynamic pricing mechanism,
we sketch the frame work which comes from our paper \cite{ChenA}.
We derive the closed form of bid-ask price of the European
contingent claim against the underlying risky asset with G-asset
price system as the discounted conditional G-expecation of the
claim, and the bid and ask prices are the viscosity solutions to
the nonlinear HJB equations.

\ \ \ \ Second, we consider the main part of this paper, i.e.,
consider the risky asset on the Knightian uncertain financial
market with the price fluctuation shows as continuous
trajectories. We propose the G-conditional full support condition
by using uncertain capacity, and the risky asset price path
satisfying the G-conditional full support condition could be
approximated by its G-consistent asset price systems. We derive
that the bid and ask prices of the European contingent claim
against such risky asset under uncertain can be expressed by
discounted of some conditional G-expectation of the claim. We give
examples, such as G-Markovian processes and the geometric
fractional G-Brownian motion \cite{ChenB}, satisfying the
G-conditional full support condition.
}
{\small \bf JEL-claasification: G10,G12,G13,D80}
\end{minipage}
\end{center}
\newpage
\section{Introduction}
The global economic crisis started from 2008 has revived an old
phillosophic idea about risk and uncertainty -- Knightian
uncertainty. Frank Knight formalized a distinction between risk
and uncertainty in his 1921 book, Risk, Uncertainty, and Profit
(\cite{Knight}). As Knight saw it, an ever-changing world brings
new opportunities for businesses to make profits, but also means
we have imperfect knowledge of future events. Therefore, according
to Knight, risk exists when an outcome can be described as draw
from a probability distribution. Uncertainty, on the other hand,
applies to situations where we cannot know all the information we
need in order to set accurate odds, i.e., we cannot know the
probability distribution for the future world. "True uncertainty,"
as Knight called it, is "not susceptible to measurement."

The study on uncertainty is still in its infancy. There is
literatures (see \cite{Bernanke83}, \cite{Bekaert},
\cite{Bloom09}, \cite{Dixit}, \cite{Gabaix}, \cite{King2010}),
\cite{Jurado}, \cite{Schwartz} and references therein) about the
macroeconomic uncertainty, which caused by uncertainty shocks,
such as, war, political and economical crisis, and terrorist
attacks, etc. For the future uncertain macroeconomic, investors
have uncertain subjective belief, which makes their consumption
and portfolio choice decisions uncertain. During a disaster an
asset's fundamental value falls by a time-varying amount. This in
turn generates time-varying risk premia and, thus, volatile asset
prices and return predictability.

The exist linear frame work can not describe the asset price in
uncertain future, uncertain risk premia, uncertain return and
uncertain volatility. How to model the asset price in uncertain
future, and pricing the uncertain risk becomes an open problem. In
this paper, we consider using G-frame work presented by Peng in
\cite{PengC}, which is a powerful and beautiful nonlinear analysis
tool, to construct the frame work to model the future risky asset
price on the Knightian uncertain financial market and pricing the
bid and ask prices of the uncertain risk.

In the first part of this paper, we define G-asset price system
(see Section 3.1) which describes the uncertain drift and
uncertain volatility of the risky asset price under uncertain. We
consider the financial market consists of the risky asset (stock)
with price fluctuation $(S_t)_{t\geq 0}$ modelled by G-asset price
system and the bond $(P_t)_{t\geq 0}$ satisfying
\begin{eqnarray}
dP_t=rP_tdt\ \ t\in [0,T],&P_0=1,
\end{eqnarray}
where $T>0$ and $r$ is short interest rate, we assume it is
constant rate without loss the technique generality. On such
financial market, the risk premium of the risky asset is uncertain
and we call it uncertain risk premium, the price of the uncertain
risk is also uncertain (see Section 3.1). We define a deflator
which implies time value and uncertain risk value. By the
technique of the G-frame work, we derive the closed forms of the
bid and ask prices of the European contingent claim against the
underlying asset with G-asset price system as conditional
G-expectations of the deflated claim.

For construct the frame work to price option against the
underlying asset with the G-asset price system, we present
G-Girsanov transform and define G-consistent dynamic pricing
mechanism, we give the expressions of the bid and ask prices of
the European contingent claim as the discounted conditional
G-expectation of claim, and the bid-ask prices are the viscosity
solutions to the nonlinear HJB equations, from which the upper
price and lower price of the European contingent claim can be
numerically computed.

The second part is our main part of this paper, we consider the
uncertain financial market with the risky asset price
$(S_t)_{t\geq 0}$ in the future expressed by a kind of continuous
trajectories which could be approximated by G-asset price systems,
and we define such G-asset price systems as G-consistent price
systems. European contingent claim against the underlying asset
with the continuous asset price path perhaps could not be priced
upper and lower prices by using G-consistent dynamic pricing
mechanism. Denote the risky price path by $S(t)$ and the portfolio
process as $\pi(t)$, and the path Riemann sum as $\sum \pi\Delta
S$. Young-Kondurar (\cite{Kondurar},\cite{Young}) Theorem tells
that the Stieltjes integral exists on certain classes of the
H$\ddot{o}$lder continuous path functions:
\begin{theorem} (Young-Kondurar Theorem))
Suppose  $\alpha$ and $\gamma$ as the H$\ddot{o}$lder exponents of
the price path $(S_t)_{t\geq 0}$ and the portfolio process path
$(\pi_t)_{t\geq 0}$, respectively, and $\gamma>1-\alpha$. Then the
integral
\begin{eqnarray}
I_t=\int_0^t\pi_tdS_t
\end{eqnarray}
exists almost surely as a limit of Riemann-Stietjes sums.
\end{theorem}
Under uncertainty, if the portfolio process and the risky asset
price path satisfying the above Theorem and the risky asset price
path has G-consistent price systems, by using G-frame work we
prove that there exists G-expectation such that the bid and ask
prices of the European contingent claim against such risky asset
have closed forms which are expressed as the discounted
conditional G-expectation of the claim. We define uncertain
capacity, by which we construct G-conditional full support
condition. The risky asset price path with the G-condition full
support condition satisfied is proved to have the G-consistent
price systems, we give examples, such as, G-Markovian process and
the geometric fractional G-Brownian motion (see \cite{ChenB}) have
the properties satisfying G-condition full support condition.

The rest of our paper is organized as follows: In Section 2 we
give notations and preliminaries for the G-frame work.  In section
3 the risky asset price model on the uncertainty future financial
market is present, which we called G-asset price system, and we
propose the G-martingale time consistent dynamic pricing mechanism
for the European contingent claim against the risky asset with
G-asset price system. In section 4 we consider uncertain risky
asset price continuous path model on uncertain financial market,
which satisfying G-conditional full support condition. And we
prove that such uncertain price model have G-consistent price
systems, and the bid-ask prices of the European contingent claim
against such uncertain risky asset can be expressed as discounted
of some conditional G-expectation of the claim. We give examples
of processes which satisfying G-conditional full support
condition.

\section{Preliminaries}
Let $\Omega$ be a given set and let $\cal{H}$ be a linear space of
real valued functions defined on $\Omega$ containing constants.
The space $\cal{H}$ is also called the space of random variables.

\begin{definition}
A sublinear expectation $\hat{E}$ is a functional
$\hat{E}:\mathcal{H}\longrightarrow R$ satisfying

(i) Monotonicity:
$$
\hat{E}[X]\geq \hat{E}[Y]\ \ \mbox{if}\ \ X\geq Y.
$$

(ii) Constant preserving:
$$
\hat{E}[c]=c\ \ \mbox{for}\ \ c\in R.
$$

(iii) Sub-additivity: For each $X,Y\in \cal{H}$,
$$
\hat{E}[X+Y]\leq \hat{E}[X]+\hat{E}[Y].
$$

(iv) Positive homogeneity:
$$
\hat{E}[\lambda X]=\lambda\hat{E}[X]\ \ \mbox{for}\ \ \lambda\geq
0.
$$
The triple $(\Omega,\mathcal{H},\hat{E})$ is called a sublinear
expectation space.
\end{definition}

In this section, we mainly consider the following type of
sublinear expectation spaces $(\Omega,\mathcal{H},\hat{E})$: if
$X_1.X_2,\dots,X_n\in\cal{H}$ then $\varphi(X_1.X_2,\dots,X_n)\in
\cal{H}$ for $\varphi\in C_{b,Lip}(R^n)$, where $C_{b,Lip}(R^n)$
denotes the linear space of functions $\phi$ satisfying
\begin{eqnarray*}
 |\phi(x)-\phi(y)|&\leq& C(1+|x|^m+|y|^m)|x-y| \mbox{ for } x,y\in
 R,\\
&& \mbox{ some } C
> 0, m\in N\mbox{ is depending on }\phi.
\end{eqnarray*}

For each fixed $p\geq 1$, we take $\mathcal{H}_0^p=\{X\in
\mathcal{H},\hat{E}[|X|^p]=0\}$ as our null space, and denote
$\mathcal{H}/\mathcal{H}_0^p$ as the quotient space. We set
$\|X\|_p:=(\hat{E}[|X|^p])^{1/p}$, and extend
$\mathcal{H}/\mathcal{H}_0^p$ to its completion
$\widehat{\cal{H}}_p$ under $\|\cdot\|_p$. Under $\|\cdot\|_p$ the
sublinear expectation $\hat{E}$ can be continuously extended to
the Banach space $(\widehat{\mathcal{H}}_p,\|\cdot\|_p)$. Without
loss generality, we denote the Banach space
$(\widehat{\mathcal{H}}_p,\|\cdot\|_p)$ as
$L^p_G(\Omega,\mathcal{H},\hat{E})$. For the G-frame work, we
refer to \cite{Peng3}, \cite{PengB}, \cite{PengA}, \cite{PengC},
\cite{PengD} and \cite{PengG}.

In this paper we assume that $\underline{\mu}, \overline{\mu},
\underline{\sigma}$ and $\overline{\sigma}$ are nonnegative
constants such that $\underline{\mu}\leq \overline{\mu}$ and
$\underline{\sigma}\leq\overline{\sigma}$.

\begin{definition} Let $X_1$ and $X_2$ be two random variables in a
sublinear expectation space $(\Omega,\mathcal{H},\hat{E})$, $X_1$
and $X_2$ are called identically distributed, denoted by
$X_1\stackrel{d}{=}X_2$ if
\begin{eqnarray*}
\hat{E}[\phi(X_1)]=\hat{E}[\phi(X_2)]& \mbox{for  } \forall\phi\in
C_{b,Lip}(R^n).
\end{eqnarray*}
\end{definition}
\begin{definition}
In a sublinear expectation space $(\Omega,\mathcal{H},\hat{E})$, a
random variable $Y$ is said to be independent of another random
variable $X$, if
\begin{eqnarray*}
\hat{E}[\phi(X,Y)]=\hat{E}[\hat{E}[\phi(x,Y)]|_{x=X}].
\end{eqnarray*}
\end{definition}
\begin{definition} (G-normal distribution) A random variable $X$
on a sublinear expectation space $(\Omega,\mathcal{H},\hat{E})$ is
called G-normal distributed if
\begin{eqnarray*}
aX+b\bar{X}=\sqrt{a^2+b^2}X&\mbox{for  } a,b\ge 0,
\end{eqnarray*}
where $\bar{X}$ is an independent copy of $X$.
\end{definition}

\begin{remark}
For a random variable $X$ on the sublinear space
$(\Omega,\mathcal{H},\hat{E})$, there are four typical parameters
to character $X$
\begin{eqnarray*}
\overline{\mu}_X=\hat{E}X,&\underline{\mu}_X=-\hat{E}[-X],\\
\overline{\sigma}_X^2=\hat{E}X^2,&\underline{\sigma}_X^2=-\hat{E}[-X^2],
\end{eqnarray*}
where $[\underline{\mu}_X,\overline{\mu}_X]$ and
$[\underline{\sigma}^2_X,\overline{\sigma}^2_X]$ describe the
uncertainty of the mean and the variance of $X$, respectively.

It is easy to check that if $X$ is G-normal distributed, then
$$
\overline{\mu}_X=\hat{E}X=\underline{\mu}_X=-\hat{E}[-X]=0,
$$
and we denote the G-normal distribution as
$N(\{0\},[\underline{\sigma}^2,\overline{\sigma}^2])$. If $X$ is
maximal distributed, then
$$
\overline{\sigma}_X^2=\hat{E}X^2=\underline{\sigma}_X^2=-\hat{E}[-X^2]=0,
$$
and we denote the maximal distribution (see \cite{PengC}) as
$N([\underline{\mu},\overline{\mu}],\{0\})$.
\end{remark}
\begin{definition}
We call $(X_t)_{t\in R}$ a d-dimensional stochastic process on a
sublinear expectation space $(\Omega,\mathcal{H},\hat{E})$, if for
each $t\in R$, $X_t$ is a d-dimensional random vector in
$\cal{H}$.
\end{definition}

\begin{definition}
Let $(X_t)_{t\in R}$ and $(Y_t)_{t\in R}$ be d-dimensional
stochastic processes defined on a sublinear expectation space
$(\Omega,\mathcal{H},\hat{E})$, for each
$\underline{t}=(t_1,t_2,\dots,t_n)\in \mathcal{T}$,
$$F_{\underline{t}}^X[\varphi]:=\hat{E}[\varphi(X_{\underline{t}})],\ \ \forall \varphi \in C_{l,Lip}(R^{n\times d})
$$
is called the finite dimensional distribution of $X_t$. $X$ and
$Y$ are said to be identically distributed, i.e., $
X\stackrel{d}{=}Y $, if
$$
F_{\underline{t}}^X[\varphi]=F_{\underline{t}}^Y[\varphi],\ \ \ \
\forall \underline{t}\in \mathcal{T}\ \ \mbox{and}\ \ \forall
\varphi\in C_{l.Lip}(R^{n\times d})
$$
where $\mathcal{T}:=\{\underline{t}=(t_1,t_2,\dots,t_n): \forall
n\in N,t_i\in R,t_i\neq t_j, 0\leq i,j\leq n,i\neq j \}$.
\end{definition}

\begin{definition}\label{Dbm1}
A process $(B_t)_{t\ge 0}$ on the sublinear expectation space
$(\Omega,\mathcal{H},\hat{E})$ is called a G-Brownian motion if
the following properties are satisfied:

(i) $B_0(\omega)=0$;

(ii) For each $t, s>0$, the increment $B_{t+s}-B_{t}$ is G-normal
distributed by
$N(\{0\},[s\underline{\sigma}^2,s\overline{\sigma}^2]$ and is
independent of $(B_{t_1},B_{t_2},\dots,B_{t_n})$, for each $n\in
N$ and $t_1, t_2,\dots,t_n\in (0, t]$;
\end{definition}

\begin{definition}
A process $(X_t)_{t\in R}$ on a sublinear expectation space
$(\Omega,\mathcal{H},\hat{E})$ is called a centered G-Gaussian
process if for each fixed $t\in R$, $X_t$ is G-normal distributed
$N(\{0\},[\underline{\sigma}_t^2,\overline{\sigma}_t^2])$, where
$0\leq \underline{\sigma}_t\leq \overline{\sigma}_t$.
\end{definition}
\begin{remark} Peng in \cite{PengC} constructs G-frame work,
which is a powerful and beautiful analysis tool for pricing
uncertain risk under uncertainty. In \cite{PengG}, Peng defines
G-Gaussian processes in a nonlinear expectation space, q-Brownian
motion under a complex-valued nonlinear expectation space, and
presents a new type of Feynman-Kac formula as the solution of a
Schr$\ddot{o}$dinger equation.
\end{remark}
In \cite{ChenB}, two-sided G-Brownian motion and fractional
G-Brownian motion are defined. The properties of the fractional
G-Brownian motion are present, such as, the similarity property
and the long rang dependent property in the sense of sublinearity,
the properties are showed in the risky asset price fluctuations in
the realistic financial market.
\begin{definition}\label{Dbm2}
A process $(B_{\frac{1}{2}}(t))_{t\in R}\in \Omega$ on the
sublinear expectation space $(\Omega,\mathcal{H},\hat{E})$ is
called a two-sided G-Brownian motion if for two independent
G-Brownian motions $(B^{(1)}_t)_{t\geq 0}$ and $(B^{(2)}_t)_{t\geq
0}$
\begin{eqnarray}
B_{\frac{1}{2}}(t)=\left\{\begin{array}{ll} B^{(1)}(t)& t\geq 0\\
B^{(2)}(-t)&t\le 0\end{array}\right.
\end{eqnarray}
\end{definition}

A family of continuous process under uncertainty which is
corresponding with the fractional Brownian motion (fBm) provided
by Kolmogorov (see \cite{Kol40} and \cite{Kol41}) and Mandelbrot
(see \cite{MvN68}) is defined as fractional G-Brownian motion
(fGBm) (see \cite{ChenB}):
\begin{definition}\label{Dfgbm}
Let $H\in (0,1)$, a centered G-Gaussian process $(B_H(t))_{t\in
R}$ on the sublinear space $(\Omega,\mathcal{H},\hat{E})$ is
called fractional G-Brownian motion with Hurst index H if

(i) $B_H(0)=0$;

(ii)
\begin{eqnarray}\label{cov}
\left\{\begin{array}{rcl}
\hat{E}[B_H(s)B_H(t)]&=&\frac{1}{2}\overline{\sigma}^2(|t|^{2H}+|s|^{2H}-|t-s|^{2H}),\
\ s,t\in R^+,\\
-\hat{E}[-B_H(s)B_H(t)]&=&\frac{1}{2}\underline{\sigma}^2(|t|^{2H}+|s|^{2H}-|t-s|^{2H}),\
\ s,t\in R^+,
\end{array}
\right.
\end{eqnarray}
we denote the fractional G-Brownian motion as fGBm.
\end{definition}
We can easily check that $(B_{\frac{1}{2}}(t))_{t\in R}$ is
G-Brownian motion, and we denote $B(t)=B_{\frac{1}{2}}(t)$. See
\cite{ChenB} for the stochastic integral with respect to fGBm.


\section{G-asset price system and G-martingale time consistent dynamic pricing mechanism}
\subsection{G-asset price model under uncertain}

The first continuous-time stochastic model for a financial asset
price appeared in the thesis of Bachelier \cite{Bachelier} (1900).
He proposed modelling the price of a stock with Brownian motion
plus a linear drift.The drawbacks of this model are that the asset
price could become negative and the relative returns are lower for
higher stock prices. Samuelson \cite{Samuelson} (1965) introduced
the more realistic model
\begin{eqnarray*}
S_t=S_0\exp{((\mu-\frac{\sigma^2}{2})+\sigma B_t^0)},
\end{eqnarray*}
which have been the foundation of financial engineering. Black and
Scholes \cite{Black} (1973) derived an explicit formula for the
price of a European call option by using the Samuelson model with
$S_0=\exp{(rt)}$ through the continuous replicate trade. Such
models exploded in popularity because of the successful option
pricing theory, as well as the simplicity of the solution of
associated optimal investment problems given by Merton
\cite{Merton} (1973).

From then on, empirical research (see \cite{Ave97}) has produced
the statistical evidence that is difficult to reconcile with the
assumption of independent and normally distributed asset returns.
Researchers have therefore attempted to build models for asset
price fluctuations that are flexible enough to cope with the
empirical deficiencies of the Black-Scholes model. In particular,
a lot of work has been devoted to relaxing the assumption of
constant volatility in the Black-Scholes model and there is a
growing literature on stochastic volatility models, see e.g., Ball
and Roma \cite{Ball} or Frey \cite{Frey} for surveys.

Knightian uncertainty can be an important factor influencing
investors' consumption and portfolio choice. Incorporating it into
asset pricing models can therefore shed light on sources of asset
return premiums and time variation in prices. In \cite{ChenA}, we
consider the asset price model on a sublinear expectation space
$(\Omega,\mathcal{H},\hat{E},\mathcal{F})$, which modelled the
stock price with uncertain drift and uncertain volatility, i.e.,
\begin{eqnarray}\label{asset}
dS_t=S_t(db_t+dB_t)
\end{eqnarray}
where $b_t+B_t$ is generalized G-Brownian motion, $b_t$ describes
uncertain drift and is distributed with
$N([\underline{\mu}t,\overline{\mu}t],\{0\})$, $B_t$ is G-Brownian
motion describes the uncertain volatility and is distributed with
$N(\{0\},[\underline{\sigma}^2t,\overline{\sigma}^2t])$,
 and $\mathcal{F}_t$ is the filtration with respect to the G-Brownian motion
$B_t$. The drift $b_t$ can be rewritten as
$$
b_t=\int_0^t\mu_tdt
$$
where $\mu_t$ is the asset return rate (\cite{ChenA})
\begin{definition} {\bf (G-asset price system)} If asset price
process $S_t$ on a sublinear space
$(\Omega,\mathcal{H},\hat{E},\mathcal{F})$ satisfying
$(\ref{asset})$, we call $(S_t,\hat{E})$ is G-asset price system.
\end{definition}
If the process $S_t$ is the asset price, and $(S_t,\hat{E})$ is
G-asset price system, we define uncertain risk premium as
\begin{definition} {\bf (Uncertain risk premium)} Assume that the
asset price is G-asset price system $(S_t,\hat{E})$, we define the
difference between return rates of the asset and bond
\begin{eqnarray}
\vartheta_t=\mu_t-r,
\end{eqnarray}
as uncertain risk premium of the asset.
\end{definition}
It is easy to prove the following Proposition (\cite{ChenA}):
\begin{proposition} The uncertain risk premium of the asset, which is G-asset price system, is uncertain and distributed by
$N([\underline{\mu}-r,\overline{\mu}-r],\{0\})$, where $r$ is the
interest rate of the bond.
\end{proposition}

\subsection{European contingent claim pricing under G-asset price system}

Consider an investor with wealth $Y_t$ in the market, who can
decide his invest portfolio and consumption at any time $t\in
[0,T]$. We denote $\pi_t$ as the amount of the wealth $Y_t$ to
invest in the stock at time $t$, and $C(t +h)-C(t)\ge 0$ as the
amount of money to withdraw for consumption during the interval
$(t, t +h],h
> 0$. We introduce the cumulative amount of consumption $C_t$ as RCLL
with $C(0) = 0$. We assume that all his decisions can only be
based on the current path information $\Omega_t$.

\begin{definition}\label{superstrategy} A self-financing superstrategy (resp.
substrategy) is a vector process $(Y,\pi,C)$ (resp. $(-Y,\pi,C)$),
where $Y$ is the wealth process, $\pi$ is the portfolio process,
and $C$ is the cumulative consumption process, such that
\begin{eqnarray}\label{eq_strategy}
\begin{array}{r}
 dY_t =rY_tdt +\pi_td B_t +\pi_t\vartheta_tdt -dC_t,\\
 \mbox{(resp. } -dY_t = -rY_tdt +\pi_td B_t+\pi_t\vartheta_tdt-dC_t\mbox{ )}
 \end{array}
 \end{eqnarray}
 where C is an increasing, right-continuous
process with $C_0 = 0$. The superstrategy (resp. substrategy) is
called feasible if the constraint of nonnegative wealth holds
$$
Y_t\geq 0,\ \ t\in[0,T].
$$
\end{definition}

We consider a European contingent claim $\xi$ written on the stock
with maturity $T$, here $\xi\in L^2_G (\Omega_T )$ is nonnegative.
We give definitions of superhedging (resp. subhedging) strategy
and ask (resp. bid) price of the claim $\xi$.
\begin{definition}
(1) A superhedging (resp. subhedging) strategy against the
European contingent claim $\xi$ is a feasible self-financing
superstrategy $(Y,\pi,C)$ (resp. substrategy $(-Y,\pi,C)$) such
that $Y_T = \xi$ (resp. $-Y_T =-\xi$). We denote by $\mathcal{H}
(\xi)$ (resp. $\mathcal{H}^{\prime}(-\xi)$) the class of
superhedging (resp. subhedging) strategies against $\xi$, and if
$\mathcal{H} (\xi)$ (resp. $\mathcal{H}^{\prime}(-\xi)$) is
nonempty, $\xi$ is called superhedgeable (resp. subhedgeable).

(2) The ask-price $X(t)$ at time $t$ of the superhedgeable claim
$\xi$ is defined as
$$
X(t)=\inf\{x\ge 0:\exists (Y_t,\pi_t,C_t)\in\mathcal{H}(\xi)\mbox{
such that } Y_t=x\},
$$
and bid-price $X^{\prime}(t)$ at time $t$ of the subhedgeable
claim $\xi$ is defined as
$$
X^{\prime}(t)=\sup\{x\ge 0:\exists
(-Y_t,\pi_t,C_t)\in\mathcal{H}^{\prime}(-\xi)\mbox{ such that }
-Y_t=-x\}.
$$
\end{definition}

Under uncertainty, the market is incomplete and the superhedging
(resp. subhedging) strategy of the claim is not unique. The
definition of the ask-price $X(t)$ implies that the ask-price
$X(t)$ is the minimum amount of risk for the buyer to superhedging
the claim, then it is coherent measure of risk of all
superstrategies against the claim for the buyer. The coherent risk
measure of all superstrategies against the claim can be regard as
the sublinear expectation of the claim, we have the following
representation of bid-ask price of the claim.

\begin{theorem}
Let $\xi\in L^2_G (\Omega_T )$ be a nonnegative European
contingent claim. There exists a superhedging (resp. subhedging)
strategy $(X,\pi,C)\in \mathcal{H}(\xi)$ (resp.
$(-X^{\prime},\pi,C)\in\mathcal{H}^{\prime}(-\xi)$) against $\xi$
such that $X_t$ (resp. $X^{\prime}_t$ ) is the ask (resp. bid)
price of the claim at time $t$.

Let$(H_s^t:s\geq t)$ be the deflator started at time $t$ and
satisfy
\begin{eqnarray}\label{eq313}
dH_s^t=-H_s^t[rds+\displaystyle\frac{\vartheta_s}{\sigma_s}d
B_s],& H_t^t=1,
\end{eqnarray}
where $\sigma_t$ is adapted process with respect to
$\mathcal{F}_t$ and $\sigma_t\in
[\underline{\sigma},\overline{\sigma}]$ (see \cite{ChenA}).

Then the ask-price against $\xi$ at time $t$ is
$$
X_t=\hat{E}[H_T^t\xi|\Omega_t],
$$
and the bid-price against $\xi$ at time $t$ is
$$
X^{\prime}_t=-\hat{E}[-H_T^t\xi|\Omega_t].
$$
\end{theorem}
{\bf Proof.} See \cite{ChenA}.
\begin{remark}
$(H_s^t:s\geq t)$ be the deflator started at time $t$ satisfying
($\ref{eq313}$), and
\begin{eqnarray}\label{eq314}
H_t=\exp\{-[\int_0^trds+\int_0^t\displaystyle\frac{\vartheta_s}{\sigma_s}d
B_s+\frac{1}{2}\int_0^t(\frac{\vartheta_s}{\sigma_s})^2d<B>_s]\}
\end{eqnarray}
which is the deflator from $0$ to $t$, and implies the time value
and the uncertain risk value.
\end{remark}

\subsection{G-Girsanov Theorem and G-martingale pricing mechanism}
In this subsection we construct the G-martingale pricing frame
work under G-asset price system. Define
\begin{eqnarray}\label{GGirsanov}
\widetilde{B}_t:=b_t+B_t-rt,
\end{eqnarray}
we have the following G-Girsanov Theorem (presented in
\cite{ChenA}, \cite{ChenB} and \cite{Humingshang})
\begin{theorem} {\bf (G-Girsanov Theorem)} Assume that $(B_t)_{t\geq 0}$ is G-Brownian motion and $b_t$ is distributed with $N([\underline{\mu}t,\overline{\mu}t],\{0\})$
on $(\Omega,\mathcal{H},\hat{E},\mathcal{F}_t)$,
and $\widetilde{B}_t$ is defined by $(\ref{GGirsanov})$, there
exists sublinear space $(\Omega,\mathcal{H},E^G,\mathcal{F}_t)$
such that $\widetilde{B}_t$ is G-Brownian motion under $E^G$, and
\begin{eqnarray}
\hat{E}[B_t^2]=E^G[\tilde{B}_t^2],&-\hat{E}[-B_t^2]=-E^G[-\tilde{B}_t^2].
\end{eqnarray}
\end{theorem}
For $t \in [0,T]$, we define G-martingale pricing mechanism as the
following conditional G-expectation $E^G_{t,T} : L^2_G (\Omega_T
)\longrightarrow L^2_G (\Omega_t )$
$$
 E^G_{t,T} [\cdot] = E^G[\cdot|\mathcal{F}_t ].
$$
The $E^G_{t,T} [\cdot]$ is a sublinear expectation, and has the
properties, such as, sub-additivity, positive constant preserving,
positive homogeneity and Chapman rule (G-Markovian Chain) (see
\cite{ChenA}), which means that $E^G_{t,T} [\cdot]$ is a time
consistent sublinear pricing mechanism (see \cite{ChenA}). By
G-martingale decomposition Theorem \cite{Song}, we can derive the
following theorem (see \cite{ChenA})
\begin{theorem}\label{th41}
Assume that $\xi = \phi(S_T)\in L^2_G (\Omega_T )$ be a
nonnegative European contingent claim, and $E^G_{t,T} [\cdot]$ be
the G-martingale pricing mechanism. The ask price and bid price
against the contingent claim $\xi$ at time t are
\begin{eqnarray}
u^a(t,S_t) = \mbox{e}^{-r(T-t)}E^G_{t,T} [\xi] \mbox{ and }
u^b(t,S_t ) = -\mbox{e}^{-r(T-t)}E^G_{t,T} [-\xi],
\end{eqnarray} respectively.

And the ask price $u^a(t,x)$ and bid price $u^b(t,x)$ against the
contingent claim $\xi$ are the viscosity solutions to the
following nonlinear HJB (proved in \cite{ChenA})
\begin{eqnarray}\label{eq521}
\partial_t u^a(t,x)+rx\partial_x u^a(t,x)+G(x^2\partial_{xx}u^a(t,x))-ru^a(t,x) = 0,& (t,x)
\in [0,T)\times R,&\\
u^a(T,x) = \phi(x).\nonumber
\end{eqnarray}
\begin{eqnarray}\label{eq522}
\partial_t u^b(t,x)+rx\partial_x u^b(t,x)-G(-x^2\partial_{xx}u^b(t,x))-ru^b(t,x) = 0,& (t,x)
\in [0,T)\times R,&\\
u^b(T,x) = \phi(x),\nonumber
 \end{eqnarray}
where the sublinear function $G(\cdot)$ is defined as follows
\begin{eqnarray}\label{eq415}
G(\alpha)=\displaystyle\frac{1}{2}(\overline{\sigma}^2\alpha^+-\underline{\sigma}^2\alpha^-),&\forall
\eta,\alpha\in R.
\end{eqnarray}

\end{theorem}

\section{G-consistent price systems and G-consistent bid-ask pricing under Uncertainty}
\subsection{G-consistent price systems}

We consider the risky asset price $S_t$ on the uncertain financial
market which shows as continuous trajectory, for example, some
path perhaps satisfying the SDE driven by fGBm (see \cite{ChenB})
or be a G-stochastic integral with average moving integrant
kernel, etc. Such process has some properties, for example, if the
price is driven by fGBm with Hurst exponent $H\in (0,1)$, the
price process is G-asset price system for $H=1/2$ and have long
range dependence if $H>1/2$. In this section we consider to study
a type of price process which has G-consistent price systems

\begin{definition} Assume that $S_t$ be a continuous price path on the sublinear space $(\Omega,\mathcal{H},\hat{E},(\mathcal{F}_t)_{t\ge 0})$,
where $\mathcal{F}_t$ is the filtration with respect to the process $S_t$.
For any $\varepsilon>0$ if there exists a G-asset price
system $(\tilde{S}_t,\tilde{E})$, such that,
\begin{eqnarray}
(1+\varepsilon)^{-1}\leq \displaystyle\frac{\tilde{S}_t}{S_t}\leq
1+\varepsilon,& \mbox{for all } t\in[0,T],
\end{eqnarray}
we call $(\tilde{S}_t,\tilde{E})$ $\varepsilon-$G-consistent price
systems, and call the continuous price process $S_t$ has
G-consistent price systems.
\end{definition}
Denote $R_{++}=(0,\infty)$, $C[u,v]$ be the set of $R-$valued
continuous functions on $[u,v]$ and $C_x[u,v]$ be all the
functions $f(t)\in C[u,v]$ with $f(u)=x$. For $x\in R_{++}$, we
denote $C_x^+[u,v]$ be the set of $R_{+}-$valued continuous
functions on $[u,v]$ starting at $x$.

\begin{definition}
On the sublinear expectation space $(\Omega,
\mathcal{H},\hat{E},(\mathcal{F}_t)_{t\ge 0})$, for $A\in
\sigma(\Omega)$ we define the capacity of $A$ as
\begin{eqnarray}
c[A]=\hat{E}[I_A],
\end{eqnarray}
where $I_A$ is the indicator function
$$
I_A:=I_A(x)=\left\{\begin{array}{ll} 1,&x\in A\\
0,&x\notin A.
\end{array}\right.
$$
\end{definition}

Let us define $S_t:= S_T$ for $t>T$, let $c(\cdot,\omega)$ be the
capacity of the $C^+[0,T]-$valued random variable $(S_{t})_{t\in
[0,T]}$,
$$
c((S_{t})_{t\in [0,T]}):=\hat{E}[I_{(S_{t})_{t\in [0,T]}}].
$$

\begin{definition}(G-Conditional Full Support) A continuous
$R_{++}-$valued process $(S_t)_{t\in [0,T]}$ satisfies
G-Conditional Full Support (GCFS) if, for all $t\in [0,T]$,
\begin{eqnarray}
\mbox{supp }c(S|_{[t,T]}|\mathcal{F}_t)= C_{S_t}^+[t,T].
\end{eqnarray}
\end{definition}
\begin{definition} (G-Strong Conditional Full Support (GSCFS))
Let $\tau$ be a stopping time of the filtration
$(\mathcal{F}_t)_{t\in [0,T]}$. Let us define $S_t:= S_T$ for
$t>T$, let $c(\cdot,\omega)$ be the capacity of the
$C^+[0,T]-$valued random variable $(S_{t})_{t\in [0,T]}$, and let
$c^{\tau}(\cdot,\omega)$ be the $\mathcal{F}_{\tau}-$conditional
capacity of the $C^+[0,T]-$valued random variable
$(S_{\tau+t})_{t\in [0,T]}$.

We say that the G-Strong Conditional Full Support Condition holds
if, for each $[0,T]-$valued stopping time $\tau$ and for almost
all $\omega\in \{\tau<T\}$, the following is true: for each path
$f\in C_{S_{\tau}(\omega)}^+[0,T-\tau(\omega)]$ and for any
$\eta>0$, $\eta-$tube around $f$ has positive
$\mathcal{F}_{\tau}-$conditional capacity, that is,
$$
c^{\tau}(B_{f,\eta}(\omega),\omega)>0,
$$
where
$$
B_{f,\eta}=\{g\in C_{S_{\tau}(\omega)}^+[0,T]:\ \
\sup_{s\in[0,T-\tau(\omega)]}|f(s)-g(s)|<\eta\}
$$
\end{definition}
With the similar argument in \cite{Schachermayer} (proof of Lemma
2.9 in \cite{Schachermayer} p.26 Appendix), we derive the
following lemma
\begin{lemma} The G-conditional full support condition (GCFS)
implies the G-strong conditional full support condition (GSCFS),
hence they are equivalent.
\end{lemma}

\begin{theorem}\label{th_GCFS} Let $(S_t)_{t\in [0,T]}$ be an adapted positive price process on
sublinear space $(\Omega,\mathcal{H},\hat{E},\mathcal{F}_t)$
satisfying GCFS condition. Then for all $\varepsilon >0$, there
exist $G$ expectation $E^G[\cdot]$ and $G-$asset price system
$((\tilde{S}_t)_{t\in [t,T]},E^G)$ in $G$ expectation space
$(\Omega,\mathcal{H},E^G,\mathcal{F}_t)$ such that
$$
|S_t-\tilde{S}_t|\leq \varepsilon,\ \ \ \ \mbox{for all } t\in
[0,T].
$$
\end{theorem}
{\bf Proof.} For $\forall\varepsilon>0$, we define the increasing
sequence of stopping times
\begin{eqnarray}
\tau_0=0,\ \ \tau_{n+1}=\inf\{t\geq
\tau_n:\displaystyle\frac{S_t}{S_{\tau_n}}\notin
((1+\varepsilon)^{-1},1+\varepsilon)\}\wedge T.
\end{eqnarray}
For $n>1$, we set
\begin{eqnarray}
R_n=\left\{\begin{array}{ll} \mbox{sign}
(S_{\tau_n}-S_{\tau_{n-1}}),&\mbox{if }\tau_n<T,\\
0,&\mbox{if }\tau_n=T.\end{array}\right.
\end{eqnarray}
Define the following Random Walk with retirement
\begin{eqnarray}
X_n=X_0(1+\varepsilon)^{\sum_{i=1}^nR_i}
\end{eqnarray}
which adapted to the discretized filtration
$\mathcal{F}_{\tau_n}$. With the similar argument in
\cite{Schachermayer} (Lemma A.1. p. 27), we have that $S_n$
satisfying GCFS implies
\begin{eqnarray}
c(R_{n+1}=z|\mathcal{F}_{\tau_n})>0,& \mbox{for }z=0,\pm 1, n>0,
\mbox{on }\{\tau_n<T\}.
\end{eqnarray}

Denote $X_{\infty}$ be terminal value of $X$, we define the
following continuous path from $X$ as
\begin{eqnarray}\label{construct}
\tilde{S}_t:=\hat{E}[X_{\infty}|\mathcal{F}_t],&t\in [0,T].
\end{eqnarray}

For $\overline{\sigma}>\underline{\sigma}\ge 0$, define a
sublinear function $G(\cdot,\cdot)$ as follows
\begin{eqnarray}\label{eq415}
G(\eta,\alpha)=(\overline{\mu}\eta^+-\underline{\mu}\eta^-)+\displaystyle\frac{1}{2}(\overline{\sigma}^2\alpha^+-\underline{\sigma}^2\alpha^-),&\forall
\eta,\alpha\in R.
\end{eqnarray}
For given $\varphi\in C_{b,lip}(R)$, we denote $u(t,x)$ as the
viscosity solution of the following G-equation (see \cite{PengC})
\begin{eqnarray}\label{eq416}
\partial_tu-G(\partial_xu,\partial_{xx}u)=0,&(t,x)\in(0,\infty)\times R,\\
u(0,x)=\varphi(x).&\nonumber
\end{eqnarray}

For $\omega\in\Omega$ consider the process
$\tilde{B}_t(\omega):=(\ln\displaystyle\frac{\tilde{S}_t}{\tilde{S}_0})(\omega)=\omega_t,
t\in [0,\infty)$, we define
 $E^G[\cdot]: \mathcal{H}\longrightarrow R$ as
\begin{eqnarray*}
E^G[\varphi(\tilde{B}_t)]=u(t,0),
\end{eqnarray*}
and for each $s,t\ge 0$ and $t_1,\cdots,t_N\in [0,t]$
\begin{eqnarray*}
E^G[\varphi(\tilde{B}_{t_1},\cdots,\tilde{B}_{t_N},\tilde{B}_{t+s}-\tilde{B}_t)]:=E^G[\psi(\tilde{B}_{t_1},\cdots,\tilde{B}_{t_N})]
\end{eqnarray*}
where
$\psi(x_1,\cdots,x_N)=E^G[\varphi(x_1,\cdots,x_N,\tilde{B}_s)]$.

For $0<t_1<t_2<\cdots<t_{i}<t_{i+1}<\cdots<t_N<+\infty$, we define
G conditional expectation with respect to $\Omega_{t_i}$ as
\begin{eqnarray*}
&&E^G[\varphi(\tilde{B}_{t_1},\tilde{B}_{t_2}-\tilde{B}_{t_1}\cdots,\tilde{B}_{t_{i+1}}-\tilde{B}_{t_{i}},\cdots,
\tilde{B}_{t_N}-\tilde{B}_{t_{N-1}})|\mathcal{F}_{t_i}]\\
&:=&\psi(\tilde{B}_{t_1},\tilde{B}_{t_2}-\tilde{B}_{t_1},\cdots,
\tilde{B}_{t_{i}}-\tilde{B}_{t_{i-1}}),
\end{eqnarray*}
where
$\psi(x_1,\cdots,x_i)=E^G[\varphi(x_1,\cdots,x_i,\tilde{B}_{t_{i+1}}-\tilde{B}_{t_{i}},
\cdots,\tilde{B}_{t_N}-\tilde{B}_{t_{N-1}})$.

We consistently define a sublinear expectation $E^G$ on
$\mathcal{H}$. Under sublinear expectation $E^G$ we define above,
the corresponding canonical process $(\tilde{B}_t)_{t\ge 0}$ is a
generalized G-Brownian motion and $(\tilde{S}_t)$ is G-asset price
system on the sublinear space $(\Omega,\mathcal{H}, E^G,
(\mathcal{F}_t)_{t\ge 0})$. We call $E^G[\cdot]$ as G-expectation
on $(\Omega,\mathcal{H},E^G[\cdot])$.

Denote $L_G^p(\Omega),p\geq 1$ as the completion of $\mathcal{H}$
under the norm $\|X\|_p=(E^G[|X|^p])^{1/p}$, and similarly we can
define $L_G^p(\Omega_t)$. The sublinear expectation $E^G[\cdot]$
can be continuously extended to the space
$(\Omega,L_G^1(\Omega))$.

For fix $t\in [0,T]$, define
$\overline{\tau}=\max\{\tau_n:\tau_n\leq t\}$ and
$\underline{\tau}=\min\{\tau_n:\tau_n>t\}$. We have
\begin{eqnarray*}
(1+\varepsilon)^{-1}\leq
\displaystyle\frac{S_t}{S_{\overline{\tau}}},
\displaystyle\frac{S_{\underline{\tau}}}{S_{\overline{\tau}}}\leq
1+\varepsilon, & \forall t\in [0,T]
\end{eqnarray*}
and therefore
\begin{eqnarray*}
(1+\varepsilon)^{-2}\leq
\displaystyle\frac{S_{\underline{\tau}}}{S_{t}}\leq
(1+\varepsilon)^2, & \forall t\in [0,T].
\end{eqnarray*}
From construction $(\ref{construct})$, for $n\ge 0$ on
$\{\tau_n<T\}$ we have $\tilde{S}_{\tau_n}=X_n, S_{\tau_n}=X_n$.
On $\{\tau_n=T\}$, we have
\begin{eqnarray*}
(1+\varepsilon)^{-1}\leq
\displaystyle\frac{\tilde{S}_{\tau_n}}{S_{\tau_n}}\leq
(1+\varepsilon), &\forall n\ge 0.
\end{eqnarray*}
Therefore, we have
\begin{eqnarray*}
\displaystyle\frac{\tilde{S}_{t}}{S_{t}}=\displaystyle\frac{E^G[\tilde{S}_{\underline{\tau}}|\mathcal{F}_t]}{S_t}
=E^G[\displaystyle\frac{\tilde{S}_{\underline{\tau}}}{S_{\underline{\tau}}}\frac{S_{\underline{\tau}}}{S_{t}}|\mathcal{F}_t]
\end{eqnarray*}
which implies
\begin{eqnarray*}
(1+\varepsilon)^{-3}\leq
\displaystyle\frac{\tilde{S}_{t}}{S_{t}}\leq (1+\varepsilon)^3.
\end{eqnarray*}
From which we complete the proof of the Theorem.$\ \ \square$

\subsection{Bid-Ask Pricing}

On the uncertain financial market if the price process
$(S_t)_{t\in[0,T]}$  of the risky asset is continuous trajectory
satisfying GCFS condition in the G-expectation space
$(\Omega,\mathcal{H},\hat{E},\mathcal{F}_t)$, we correct the
definition $(\ref{eq_strategy})$ of the self-finance superheding
(resp. subheding) strategy $(Y,\theta,C)$ (resp. $(-Y,\theta,C)$)
as following
\begin{eqnarray}\label{self_finance}
\begin{array}{r}
Y_t =\int_0^t(Y_s-\theta_sS_s)rds +\int_0^t\theta_sdS_s -C_t,\\
(\mbox{resp. }-Y_t =\int_0^t(-Y_s-\theta_s)rds
+\int_0^t\theta_sS_sdS_s -C_t,
\end{array}
\end{eqnarray}
where $\theta_t=\displaystyle\frac{\pi_t}{S_t}$, the integral
$\int_0^t\theta_sdS_s$ is in a pointwise Riemann-stietjes sense,
and $C_t$ is a right continuous, nondecreasing cost process with
$C_0=0$, i.e.,
\begin{eqnarray}
\begin{array}{r}
dY_t =(Y_t-\theta_tS_t)rdt +\theta_tdS_t -dC_t.\\
(\mbox{resp. }-dY_t =(-Y_t-\theta_tS_t)rdt +\theta_tdS_t -dC_t.
\end{array}
\end{eqnarray}

A superhedging (resp. subhedging) strategy against the European
contingent claim $\xi$ is a feasible self-financing superstrategy
$(Y,\theta,C)$ (resp. substrategy $(-Y,\theta,C)$) such that $Y_T
= \xi$ (resp. $-Y_T =-\xi$). We denote by $\mathcal{H}_c (\xi)$
(resp. $\mathcal{H}_c^{\prime}(-\xi)$) the class of superhedging
(resp. subhedging) strategies against $\xi$, and if $\mathcal{H}_c
(\xi)$ (resp. $\mathcal{H}_c^{\prime}(-\xi)$) is nonempty, $\xi$
is called superhedgeable (resp. subhedgeable).

\begin{definition}
The ask-price $X_{c}(t)$ at time $t$ of the superhedgeable claim
$\xi$ is defined as
$$
X_{c}(t)=\inf\{x\ge 0:\exists
(Y_t,\theta_t,C_t)\in\mathcal{H}_c(\xi)\mbox{ such that } Y_t=x\},
$$
and the bid-price $X_{c}^{\prime}(t)$ at time $t$ of the
subhedgeable claim $\xi$ is defined as
$$
X_{c}^{\prime}(t)=\sup\{x\ge 0:\exists
(-Y_t,\theta_t,C_t)\in\mathcal{H}_c^{\prime}(-\xi)\mbox{ such that
} -Y_t=-x\}.
$$
\end{definition}
\begin{theorem}
Let $(S_t)_{t\in [0,T]}$ be an $R_{++}-$valued continuous process
on the sublinear expectation space
$(\Omega,\mathcal{H},\hat{E},\mathcal{F}_t)$ satisfying the
G-conditional full support assumption (GCFS) and the nonnegative
European contingent claim $\xi=g(S_T)\in L_G^2(\Omega_T)$.

Then there exists G-expectation $E^G$, such that the ask and bid
prices of the European contingent claim $g(S_T)$ at time $t$ are
given by
\begin{eqnarray}
X_c(t)=e^{-r(T-t)}E^G[g(S_t)|\mathcal{F}_t]&X_c^{\prime}(t)=-e^{-r(T-t)}E^G[-g(S_T)|\mathcal{F}_t]
\end{eqnarray}
respectively.
\end{theorem}
{\bf Proof.} By Theorem $\ref{th_GCFS}$, for $\forall \varepsilon
>0 $, there exist $\varepsilon$-G-consistent price systems $(\tilde{S}_t,\tilde{E})$
satisfying
$$
(1+\varepsilon)^{-1}\leq \displaystyle\frac{\tilde{S}_t}{S_t}\leq
1+\varepsilon.
$$
For small enough $\varepsilon$, we denote the family of the
$\varepsilon$-G-consistent price systems as
$$
\mathcal{Z}_{\varepsilon}:=\{(\tilde{S}_t,\tilde{E}): \tilde{S}_t
\mbox{ is a G-asset price, }1-\varepsilon\leq
\displaystyle\frac{\tilde{S}_t}{S_t}\leq 1+\varepsilon,t\in[0,T]
\}.
$$
For fix $\varepsilon$, with the corresponding
$\varepsilon$-G-consistent price system $(\tilde{S}_t,\tilde{E})$
there exist adapt processes $\delta_{t,1}$ and $\delta_{t,2}$
satisfying
\begin{eqnarray}\label{numdelta}
\begin{array}{l}
\delta_{t,1}\in [-\varepsilon,\varepsilon],\ \ \ \ \ \delta_{t,2}\in [-2\varepsilon,2\varepsilon]\\
\delta_{t,1}\delta_{t,2}\neq 0,\ \ \mbox{a.s.}
\end{array}
\end{eqnarray}
such that
\begin{eqnarray}\label{StoS}
S_t=(1+\delta_{t,1})\tilde{S}_t,& dS_t=2\delta_{t,2}d\tilde{S}_t,\
\ \ \forall t\in [0,T].
\end{eqnarray}
Denote $\tilde{B}_t$ as the G-Brownian motion on the sublinear
expectation space $(\Omega,\mathcal{H},\tilde{E},\mathcal{F}_t)$
distributed with
$N(\{0\},[\underline{\sigma}^2t,\overline{\sigma}^2t])$, from the
construction of the process $\tilde{S}_t$ in Theorem
$\ref{th_GCFS}$, process $\tilde{S}$ is a G-asset price process
\begin{eqnarray*}
d\tilde{S}_t=\tilde{S}_t(d\tilde{b}_t+d\tilde{B}_t),
\end{eqnarray*}
where $\tilde{b}_t$ is distributed with
$N([\underline{\mu}t,\overline{\mu}t],\{0\})$ in
$(\Omega,\mathcal{H},\tilde{E},\mathcal{F}_t)$.

By G-Girsanov transform (\cite{ChenA}, \cite{ChenB} and
\cite{Humingshang}) there exists a G-expectation space
$(\Omega,\mathcal{H},E^G,\mathcal{F}_t)$ such that
\begin{eqnarray}\label{Ggirsavov}
B_t^G:=\tilde{B}_t+\tilde{b}_t-\int_0^t\displaystyle\frac{1+\delta_{t,1}}{2\delta_{t,2}}rdt,&
t\geq 0
\end{eqnarray}
is a G-Brownian motion in
$(\Omega,\mathcal{H},E^G,\mathcal{F}_t)$.

Define process $X_t$ as follows
\begin{eqnarray}
X_t:=e^{-r(T-t)}E^G[g(S_T)|\mathcal{F}_t],
\end{eqnarray}
we have that $e^{-rt}X_t=E^G[e^{-rT}g(S_T)|\mathcal{F}_t]$ is a
G-martingale in $(\Omega,\mathcal{H},E^G,\mathcal{F}_t)$, by the
G-martingale representation Theorem \cite{Song}
\begin{eqnarray}
e^{-rt}X_t=E^G[e^{-rT}g(S_T)]+\int_0^t\beta_sdB_t^G-K_t,
\end{eqnarray}
where $\beta_t\in L_G^1[0,T]$, $K_t$ is a continuous, increasing
process with $K_0=0$, and $\{K_t\}_{t\in [0,T]}$ is a
G-martingale. Define
\begin{eqnarray*}
\theta_t:=\displaystyle\frac{e^{rt}\beta_t}{2\delta_{t,2}\tilde{S}_t},&C_t=\int_0^te^{rs}K_s
ds
\end{eqnarray*} we derive that
\begin{eqnarray*}
dX_t=rX_tdt+2\theta_t\delta_{t,2}\tilde{S}_tdB_t^G-dC_t
\end{eqnarray*}
which implies that
\begin{eqnarray*}
X_t=g(S_T)+\int_t^T(X_s-\theta_sS_s)rds+\int_t^T\theta_sdS_s-(C_T-C_t).
\end{eqnarray*}
Thus, we prove that
$(E^{G}[e^{-r(T-t)}g(S_T)]|\mathcal{F}_t],\theta_t,C_t)\in
\mathcal{H}_c(\xi)$ is a superhedging strategy against the claim
$\xi=g(S_T)$.

For given any superhedging strategy
$(\bar{X}_t,\bar{\theta}_t,\bar{C}_t)\in \mathcal{H}_c(\xi))$
against the claim $\xi=g(S_T)$
\begin{eqnarray*}
\bar{X}_t=g(S_T)+\int_t^T(\bar{X}_s-\bar{\theta}_sS_s)rds+\int_t^T\bar{\theta}_sdS_s-(\bar{C}_T-\bar{C}_t).
\end{eqnarray*}
From $(\ref{numdelta})$, $(\ref{StoS})$ and G-Girsanov transform
$(\ref{Ggirsavov})$, the above equation can be rewritten as
\begin{eqnarray*}
d(e^{-rt}\bar{X}_t)=2e^{-rt}\bar{\theta}_t\delta_{t,2}\tilde{S}_tdB_t^G-e^{-rt}d\bar{C}_t
\end{eqnarray*}
\begin{eqnarray*}
e^{-rt}\bar{X}_t=e^{-rT}g(S_{T})-\int_t^T2e^{-rt}\bar{\theta}_s\delta_{s,2}\tilde{S}_sdB_s^G+\int_t^Te^{-rt}d\bar{C}_t
\end{eqnarray*}
take G conditional expectation with respect to $\mathcal{F}_t$,
notice that the cost function $\bar{C}_t$ is nonnegative and
nondecreasing process, we have that $\bar{X}_t\geq
e^{-r(T-t)}E^G[g(S_{T})|\mathcal{F}_t]=X_t$, which prove that
$X_c(t)=e^{-r(T-t)}E^G[g(S_{T})|\mathcal{F}_t]$.

Similarly, we can prove
$X_c^{\prime}(t)=-e^{-r(T-t)}E^G[-g(S_{T})|\mathcal{F}_t]$.$\ \
\square$

\subsection{Examples}
\begin{example} {\bf G-Markovian processes} \end{example}
Denote $\hat{E}_t[\cdot]:=\hat{E}[\cdot|\mathcal{F}_t]$, we
consider continuous nonnegative G-Markovian processes $(S_t)_{t\in
[0,T]}$ in $(\Omega,\mathcal{H},\hat{E},\mathcal{F}_t)$, defined
by
\begin{eqnarray}
\hat{E}[\phi(S_s)|\mathcal{F}_t]=\phi(S_t),& s\geq t,\ \
\forall\phi\in C_{b,Lip}(R).
\end{eqnarray}

The G-Markovian property implies the GCFS
\begin{eqnarray}
\mbox{supp } c(S|_{[v,T]}|\mathcal{F}_v)=\mbox{supp }
c(S|_{[v,T]}|S_v)=C_{S_v}^+[v,T], &0\leq v\leq T.
\end{eqnarray}

\begin{example}{\bf Processes Driven by Fractional G-Brownian
Motion}\end{example} Denote $B_t^H$ as fractional G-Brownian
motion with Hurst index $H\in (0,1)$, which is defined in
\cite{ChenB} as a centered G-Gaussian process with stationary
increment in the sense of sublinear
\begin{eqnarray*}
\overline{R}(t,s):=\hat{E}[B_t^HB_s^H]=\displaystyle\frac{\overline{\sigma}^2}{2}(t^{2H}+s^{2H}+|t-s|^{2H}),\\
\underline{R}(t,s):=-\hat{E}[-B_t^HB_s^H]=\displaystyle\frac{\underline{\sigma}^2}{2}(t^{2H}+s^{2H}+|t-s|^{2H}).
\end{eqnarray*}
The moving representation of the fractional G-Brownian motion (see
Theorem 1 in \cite{ChenB}) is
\begin{eqnarray}\label{Efgbm}
B_H(t,\omega)=C_H^w\int_{R}[(t-s)_+^{H-1/2}-(-s)_+^{H-1/2}]dB(s,\omega),
\end{eqnarray}
where $C_H^w=\displaystyle\frac{(2H\sin{\pi H
}\Gamma(2H))^{1/2}}{\Gamma(H+1/2)}$ and $(B_t)_{t\in R}$ is a
two-sided G-Brownian motion.

Denote
$R_H(t,s)=\displaystyle\frac{1}{2}(t^{2H}+s^{2H}+|t-s|^{2H})$,
then there exists square-integrable kernel $K_H(t,s)$ such that
\begin{eqnarray}
B_t^H=\int_0^tK_H(t,s)dB_s,
\end{eqnarray}
where $(B_t)_{t\in [0,T]}$ is G-Brownian motion. $(B_t)_{t\in
[0,T]}$ generate the same filtration as $(B_t^H)_{t\in [0,T]}$,
and $K_H(t,s)$ is as following
\begin{eqnarray}
K_H(t,s)=\left\{\begin{array}{ll}
C_{H,1}[(\frac{t}{s})^{H-\frac{1}{2}}(t-s)^{H-\frac{1}{2}}-(H-\frac{1}{2})s^{\frac{1}{2}-H}\int_s^tu^{H-\frac{3}{2}}(u-s)^{H-\frac{1}{2}}du],&H<
\frac{1}{2}\\
C_{H,2}s^{\frac{1}{2}-H}\int_s^t(u-s)^{H-\frac{3}{2}}u^{H-\frac{1}{2}}du,&t>s,\
H\geq {1}{2}
\end{array}\right.
\end{eqnarray}
where
\begin{eqnarray*}
C_{H,1}&=&(\frac{2H}{(1-2H)\beta(1-2H,H+1/2)})^{1/2}\\
C_{H,2}&=&(\frac{H(2H-1)}{\beta(2-2H,H-\frac{1}{2})})^{1/2}
\end{eqnarray*}
and $\beta$ denotes the Beta function.

It is easy to check that for any $v\in [0,T]$, the process
$(B_t^H)_{t\in [v,T]}$ is G-Gaussian, conditionally on
$\mathcal{F}_t$ in the sense of finite-dimensional distributions
(see \cite{PengG}), and its conditional G-expectation and
conditional increment function in the sense of sublinear are given
\begin{eqnarray*}
\hat{E}[B_t^H|\mathcal{F}_v]=\int_0^vK_H(t,s)dB_s,&t\geq v,\\
\hat{E}[B_t^HB_s^H|\mathcal{F}_v]=\overline{\sigma}^2\int_v^{t\wedge
s}K_H(t,u)K_H(s,u)du,&t,s\geq v,\\
-\hat{E}[-B_t^HB_s^H|\mathcal{F}_v]=\underline{\sigma}^2\int_v^{t\wedge
s}K_H(t,u)K_H(s,u)du,&t,s\geq v.
\end{eqnarray*}
Then the law of $(B_t^H)_{t\in [v,T]}$ conditional on
$\mathcal{F}_v$ is identical with the law of
$\hat{E}[B_t^H|\mathcal{F}_v]+X_t$, where $(X_t)_{t\in [v,T]}$ is
still a fractional G-Browian motion start from $v$, i.e., is a
centered G-Gaussian process with continuous path on $[v,T]$. With
a similar argument as above, $X_t=\int_v^tK_H(t,s)dB_s$. We just
need to prove that the centered G-Gaussian process $(X_t)_{t\in
[v,T]}$ has full support as follows
\begin{eqnarray}
\mbox{supp }c(X|_{[v,T]}|\mathcal{F}_v)=C_0[v,T],
\end{eqnarray}
where $c(X|_{[v,T]}|\mathcal{F}_v)$ is the capacity on
$X|_{[v,T]}|\mathcal{F}_v$. By using the properties of the
capacity, we have the similar result as Theorem 3 in
\cite{kallianpur}
\begin{theorem}
For the centered G-Gaussian process described by
\begin{eqnarray*}
X_t=\int_v^tK_H(t,s)dB_s,
\end{eqnarray*}
the support of $(X_t)_{t\in [v,T]}$ satisfying
\begin{eqnarray*}
\mbox{supp }c(X|_{[v,T]}|\mathcal{F}_v)=\overline{H(K_H)},
\end{eqnarray*}
where $H(K_H)$ is reproducing kernel Hilbert space define by
\begin{eqnarray*}
H(K_H):=\{f(t)\in C_0([v,T],R):f(t)=\int_v^TK_H(t,s)g(s)ds,\ \
\mbox{for some }g\in L^2[v,T] \}.
\end{eqnarray*}
\end{theorem}
Define the kernel operator $K_H$ as
\begin{eqnarray*}
(K_Hg)(t):=\int_0^tK_H(t,s)g(s)ds,&g\in L^2[0,T],t\in[0,T],
\end{eqnarray*}
then for $g\in C_0[v,T]$, $K_H:C_0[v,T]\longrightarrow C_0[v,T]$
is continuous and has  a dense range (see \cite{Schachermayer}),
and $H(K_H)$ is norm-dense in $C_0([v,T],R)$, thus we have
\begin{theorem}\label{fGBmGCFS} Processes $(S_t)_{t\in [0,T]}$ in
$(\Omega,\mathcal{H},\hat{E},\mathcal{F}_t)$ driven by fractional
G-Brownian motion $B_t^H$
\begin{eqnarray*}
dS_t=S_t(b(t)dt+dB_t^H)
\end{eqnarray*}
where $b(t)$ is a deterministic continuous function, $H\in (0,1)$,
then $(S_t)_{t\in [0,T]}$ satisfies the G-conditional full support
condition (GCFS).
\end{theorem}



\end{document}